\documentclass[final,11pt]{elsarticle}
\usepackage{amsmath}
\usepackage{amsfonts}
\usepackage{amsthm}

\newtheorem{theorem}{Theorem}[section]
\newtheorem{lemma}[theorem]{Lemma}

\journal{}

\begin{document}

\begin{frontmatter}

\title{
Large-time behavior of the weak solution to 3D Navier-Stokes
equations.}

\author{A. G. Ramm}

\address{Mathematics Department, Kansas State University, Manhattan,\\
KS 66506, USA; email: ramm@math.ksu.edu}

\begin{abstract}

The weak solution to the Navier-Stokes equations in a bounded domain
$D \subset \mathbb{R}^3$ with a smooth boundary is proved to be
unique provided that it satisfies an additional requirement. This
solution exists for all $t \geq 0$. In a bounded domain $D$ the
solution decays exponentially fast as $t\to \infty$ if the force
term decays at a suitable rate.

\end{abstract}


\end{frontmatter}

\section{Introduction} \label{Introduction} Consider the
problem
\begin{equation}\label{eq:1} v' + (v, \nabla)v = -\nabla p +
\nu\Delta v + f \, \text{ in } \, D, \quad \nabla \cdot v =
0,\end{equation}
\begin{equation}\label{eq:2} v(x, 0) = v^{0}(x); \quad
v|_{S} = 0.
\end{equation}
Here $v = \displaystyle(v_m)_{m = 1}^{3}$ is a vector function, $v'
= \frac{dv}{dt}$, $D \subset \mathbb{R}^3$ is a bounded domain with
a smooth boundary $S$, $\nu = \text{const} > 0$ is the kinematic
viscosity coefficient, $v^{0}$ and $f$ are given functions, $v$ and
$p$ are to be found. We assume throughout that $v^{0}(x) \in
\mathring{H}^1(D)$, $\nabla \cdot v^{0}=0$, and $f \in L^2([0, T);
H^1(D))$ for any $T<\infty$. We also assume that $f$ decays fast as
$t\to \infty$. Precise assumptions will be formulated in Section 2,
in the proof of Lemma 2.1.

We use the standard  notations: $\mathring{H}^1(D)$ is the closure
of vector-functions $C^\infty_0(D)$ in the norm of the Sobolev space
$H^1(D)$;
 $V$ is the closure in $H^1(D)$ of the subset of $C^\infty_0(D)$
consisting of solenoidal vector fields, $\nabla \cdot v=0$; $(u,v)$
is the inner product in $H:=L^2(D)$ of two vector functions in
$\mathbb{R}^3$,
$$|u|^2:=(u,u),\quad ((u,v)):=(\nabla u, \nabla v),\quad ||u||^2:=((u,u)).$$

 {\bf Definition 1.} {\it A weak solution to (\ref{eq:1})-(\ref{eq:2}) is a vector function  $v \in
W:=L^2([0, T); V)$ satisfying the relation}
\begin{equation}\label{eq:3}
(v',\eta)+((v\cdot \nabla)v,\eta) +\nu ((v,\eta))=(f,\eta) \quad \forall \eta \in W.
\end{equation}
One proves that  $(v',\eta)\in L^1([0,T))$ if $v\in V$. Indeed,
$((v,\eta))\in L^2([0,T))$ because $v\in L^2([0,T); V)$ and $\eta\in
V$, so $\eta\in L^\infty([0,T); V)$. An integration by parts and
H\"older's inequality yield
$$|((v\cdot
\nabla)v,\eta)|=|-(v v, \nabla \eta)|\le ||v||^2_{L^4(D)}||\eta||.$$
Here $(v v, \nabla \eta):=(v_jv_m,\eta_{m,j})$, over the repeated
indices summation is understood, $v_m$ is the $m-$th Cartesian
component of the vector function $v$, $\eta_{m,j}:=\frac {\partial
\eta_m}{\partial x_j}$. We use below the multiplicative inequality
$$||v||^2_{L^4(D)}\le c|v|^{1/2}||v||^{3/2}, \quad c=const>0,$$
(see [5]), and the Young's inequality
$$ab\le \frac{\epsilon^pa^p}{p}+\frac{\epsilon^{-q}b^q}{q},\,\,\,
\forall \epsilon>0;\quad \frac 1 p+\frac 1 q=1,\,\,a,b>0.$$
 By $c$ we denote throughout this paper {\it
various} positive {\it time independent}  constants. Using the
Young's inequality with $\epsilon=1$ and $p=4$, one gets
$|v|^{1/2}||v||^{3/2}\le \frac {|v|^{2}}{4}+ \frac{3||v||^{2}}{4}$.
Since $|v|^{2}\in L^1([0,T))$ and $||v||^{2}\in L^1([0,T))$, it
follows from equation (\ref{eq:3}) that $(v',\eta)\in L^1([0,T))$
because all other terms in this equation are in $L^1([0,T))$. If
(\ref{eq:3}) holds for all $\eta\in W$, then it holds for all $\eta
\in V$, and vice versa, because the set of functions
$\eta(x)\phi(t)$ for $\eta\in V$ and $\phi\in L^2([0,T))$ is dense
in $W$. Thus, relation (\ref{eq:3}) is well defined for $v\in W$ and
$\eta\in V$.

The questions of interest are: a) Is the weak solution unique? b)
Does it exist globally, that is, for all $t\ge 0$? c)  How does it
 behave as $t\to \infty$?
 d) Is it smooth if the data are smooth?  e) Does the smooth solution to
(\ref{eq:1})-(\ref{eq:2}) exist globally? f) Does its smoothness
improves if the smoothness of the data improves?

These questions were discussed in several books and many papers, see
[1]-[7] and references therein. Existence of the weak solutions was
proved in [5]-[7], but its uniqueness was not proved, and, for a long 
time, it has been an open problem to prove uniqueness of the weak 
solution.    
Local existence of the smooth solution and its uniqueness was proved in 
the cited books. The
smoothness properties of the weak solution are improving locally if
the smoothness of the data improves. Methods for proving this are
developed in [3],[5]-[7], where theorems of this type can be found.

Let $W_1\subset W$ denote a subset of $W$ that consists of the
elements $v$ such that
\begin{equation}\label{eq:W}
||v(t)||\le c.
\end{equation}
By $c$ here and below various positive constants, independent of $t$, 
are denoted.

 The basic
results of this paper include the proof of the uniqueness of the
weak solution $v\in W_1$ and the decay estimates for the weak
solutions as $t\to \infty$.
\begin{theorem}\label{theorem1} Problem (\ref{eq:3}) has at most one
solution $v\in W_1$.
\end{theorem}
\begin{theorem}\label{theorem2} A weak solution in $W$ exists globally and 
decays exponentially
fast as $t\to \infty$ provided that the force term decays
sufficiently fast.
\end{theorem}
The decay estimates for the solution of problem (\ref{eq:3}) are
given in Lemma 2.1.

The known sufficient condition for the uniqueness of the weak solution is 
the Serrin's condition (see [7], p.276). If $v\in W_1$, or 
inequality (\ref{eq:A}) (see below) holds,  then 
the Serrin's condition holds. Therefore, the result of Theorem 1.1 can 
be obtained as a 
consequence of the Serrin's uniqueness result (cf Theorem 1.5.1  on p.276 
in [7]).  Our proof is based on the estimates given in Lemma 2.1, it 
is short, and it uses minimal background.
   
The exponential decay of solutions to Navier-Stokes equations has been
discussed in [7], p. 337, for the domains for which the 
Poincar$\acute{e}$
inequality holds. Our proof is different and shorter. 
Moreover, our estimates are valid, in contrast to the ones in [7],
also in the case when the data do not decay exponentially fast as $t\to 
\infty$,
see the last statement in Lemma 2.1.
We derive estimates using a nonlinear differential inequality. 
The presentation
in this paper is essentially self-contained.

In section 2 a proof of Theorem 1.1 is given and estimates of the
solution as $t\to \infty$ are derived in Lemma 2.1. In Section 3 a
proof of the existence part of Theorem 1.2 is given. In Section 4
the case of unbounded domain is discussed.

\section{Proof of Theorem 1.1} {\bf 2.1.} {\bf Some inequalities.}

If $D \subset \mathbb{R}^3$ is a bounded domain then
$\mathring{H}^1(D) \subset L^q(D)$, $q < 6$, and
\begin{equation}\label{eq:Z}
||v||^2_{L^4(D)} \leq c||v||^{1/2}_{L^2(D)}||\nabla v||^{3/2}_{L^2(D)} \leq
\epsilon||v||^2 + \frac{c}{4\epsilon}|v|^2,
\,\,\forall \epsilon > 0.
\end{equation}
Similar inequalities hold also if $D = \mathbb{R}^3$. For example,
\begin{equation}\label{eq:Y}
||v||^2_{L^4(\mathbb{R}^3)}=2||v||^{1/2}_{L^2(\mathbb{R}^3)}||
\nabla v||^{3/2}_{L^2(\mathbb{R}^3)}\le \epsilon ||\nabla
v||^2_{L^2(\mathbb{R}^3)}+ c(\epsilon)||v||^2_{L^2(\mathbb{R}^3)},
\end{equation}
where $\epsilon>0$ can be arbitrarily small, and  the Young's inequality 
was used.

Let $\eta = v$ in (\ref{eq:3}) and get
\begin{equation}\label{eq:4}
|v|^2+2\nu\int_0^t||v||^2ds=|v^0|^2+2\int_0^t(f,v)ds\le c+\int_0^t|f||v|ds,
\end{equation} where
$((v \cdot \nabla)v, v) = 0$ if $\nabla \cdot v = 0$ and $v|_S = 0$.
If $\int_0^\infty |f(s)|ds<\infty$, then Lemma 2.1 (see formula
(\ref{eq:11}) below) yields the estimate $\sup_{t\in [0,T)}|v(t)|\le c$. 
This estimate and
inequality (\ref{eq:5}) imply that $\sup_{t\in 
[0,T)}\int_0^t||v(s)||^2ds\le c$.

 {\bf 2.2.} {\bf Large-time behavior of solutions.}

 Let us derive
some estimates from (\ref{eq:4}). Denote $$g(t):= |v|^2,\quad
G(t):=||v||^2, \quad b(t):=|f|.$$ Differentiate (\ref{eq:4}) with
respect to $t$ and get
\begin{equation}\label{eq:5} g'(t) + 2\nu G(t) =
2(f,v) \leq 2b(t)g^{1/2}(t).
  \end{equation}
As was mentioned below formula  (\ref{eq:3}),  the derivative $v'$ 
exists in the sense that for all $\eta\in W$ one has  
$(v',\eta)\in L^1([0,T))$ if $v\in V$. 
 Let us assume that
\begin{equation}\label{eq:6} \lim_{t \to \infty}b(t) = 0,\quad \lim_{t \to
\infty}\frac{b'(t)}{b(t)} = 0. \end{equation} If $D$ is a finite domain
then \begin{equation}\label{eq:7} || v||^2 \geq c_D|v|^2, \quad v
\in \mathring{H}^1(D),\quad c_D=const>0. \end{equation}
 Thus, $G\ge c_D g$, and inequality (\ref{eq:5}) implies
\begin{equation}\label{eq:8} g' + 2 \gamma g \leq
2b(t)g^{1/2},\quad \gamma:= \nu c_D >0,\quad t \geq 0.
\end{equation}
\begin{lemma}\label{lemma1}
Assume that $g\ge 0$ and inequality (\ref{eq:8}) holds. Then
\begin{equation}\label{eq:9} g^{1/2}(t) \leq e^{-\gamma
t}g^{1/2}(0) + \frac{1}{2}\int_0^te^{-\gamma(t - s)}b(s)ds,
\end{equation}
and
\begin{equation}\label{eq:10} g(t) \leq 2e^{-2\gamma
t}g(0) + \frac{1}{2}\bigg(\int_0^te^{-\gamma(t - s)}b(s) ds\bigg)^2.
\end{equation}
Assume that $b(t) > 0$ and conditions (\ref{eq:6}) hold. Then
\begin{equation}\label{eq:11} \lim_{t \to
\infty}\frac{\int_0^te^{-\gamma(t - s)}b(s)ds}{b(t)} =
\frac{1}{\gamma},\quad \gamma > 0. \end{equation}
\end{lemma}
{\it Proof of Lemma 2.1}. Let $h(t) = g(t)e^{2\gamma t}$. Then
$h(0)=g(0)$. Equation (\ref{eq:8}) implies $h' \leq 2b(t)e^{\gamma
t}h^{1/2}$. So, $$h^{1/2}(t) \leq h^{1/2}(0) + \int_0^tb(s)e^{\gamma
s}ds,$$ and (\ref{eq:9}) follows. Inequality (\ref{eq:10}) follows
from (\ref{eq:9}) since $(a + b)^2 \leq 2(a^2 + b^2)$.  Relation
(\ref{eq:11}) follows from the L'Hospital rule and conditions 
(\ref{eq:6}).

Lemma 2.1 is proved. \hfill $\Box$

{\bf Remark 1.} If $b(t) = 0$ for $t > t_0$, then (\ref{eq:10})
yields
$$g(t) \leq 2e^{-2\gamma t}g(0) + \frac{e^{-2\gamma
t}}{2}\bigg(\frac{e^{\gamma t_0} - 1}{\gamma}\bigg)^2= O(e^{-2\gamma
t}).$$  If $b(t) = O(e^{-kt})$ and $k < \gamma$, then $g(t) \leq
O(e^{-2kt})$. If $k > \gamma$, then $g(t) \leq O(e^{-2\gamma
t})$.  From (\ref{eq:6}) and (\ref{eq:8})--(\ref{eq:11}) one gets
$g'(t) \leq O(e^{-\gamma t}+b(t)).$ If $b(t)=O(e^{-\gamma t})$, then
\begin{equation}\label{eq:12}g'(t) \leq O(e^{-\gamma t}).
 \end{equation}

Estimates in Lemma 2.1 and Remark 1 prove the part of Theorem 1.2.
that deals with large-time behavior of the solution to (\ref{eq:3}).
The last statement of Lemma 2.1 allows one to prove decay estimates 
when the decay of the data $f$, as $t\to \infty$ is much slower than 
an exponential. Remember that $b(t)=|f(t)|$ is defined by the data.
Conditions  (\ref{eq:6}) and the last statement of Lemma 2.1   
allow one to estimate the rate of decay of the integral in formula 
(\ref{eq:9}). Conditions  (\ref{eq:6}) hold, for example, if 
$c_1t^{-a_1}\le b(t)\le ct^{-a}$ and $|b'(t)|\le ct^{-a-1}$,
where $0<a_1\le a$, so the decay of the data is much slower than 
an exponential. This case is not covered by the results in [7]. 

\hfill $\Box$

{\bf 2.3.} {\bf Proof of the uniqueness of the solution to
(\ref{eq:3}) in the space $W_1$.}

Suppose that $v, w \in W$
 solve (\ref{eq:3}). Let $u = v - w.$ Subtract
(\ref{eq:3}) with $w$ in place of $v$ from (\ref{eq:3}) and get
\begin{equation}\label{eq:13} (u',\eta)+\nu((u,\eta))+((u\cdot \nabla)v,\eta)+((w\cdot\nabla)u,\eta) = 0,
\,\,\,\forall \eta \in W.
\end{equation}
Take $\eta = u$ and use the relation $((w\cdot\nabla)u,u)=0$ which
holds for $u,w\in W$. Denote $h:=|u|^2$, $H:=||u||^2$. Then relation
(\ref{eq:13}) and H\"older's inequality yield
\begin{equation}\label{eq:14}
h'(t)+2\nu H(t)\le ||v(t)||||u||^2_{L^4(D)}\le c||u||^2_{L^4(D)},
\end{equation}
where the assumption $||v(t)||\le c$ was used. From (\ref{eq:14})
one gets
\begin{equation}\label{eq:14a}
h(t)+2\nu\int_0^tH(s)ds\le c\int_0^t||u||^2_{L^4(D)}ds.
\end{equation}
Using inequality (\ref{eq:Z}) one gets
\begin{equation}\label{eq:14b}
||u||^2_{L^4(D)}\le c|u|^{1/2}||u||^{3/2}\le \nu H+c(\nu)h,
\end{equation}
where the Young's inequality was used.

Since $H\ge 0$,  inequalities (\ref{eq:14}) and (\ref{eq:14b}) yield
$h'\le ch$, and $h(0)=0$ by the assumption. Therefore
\begin{equation}\label{eq:15}
h(t)\le c\int_0^t h(s)ds, \quad h(0)=0.
\end{equation}
This implies that $h=0\,\, \forall t\ge 0$.  The assumption
 (\ref{eq:W}) was crucial for the proof. Theorem 1.1 is proved.
\hfill $\Box$

{\bf Remark 2.} A slight variation of the above argument shows that
the additional assumption (\ref{eq:W}) can be replaced by the
assumption
\begin{equation}\label{eq:A}
\int_0^t ||v(s)||^4ds\le c.
\end{equation}
Recall that $c>0$ is independent of $t$.

\section{ Global existence of the weak solution}
In this Section the existence part of Theorem 1.2 is proved. The
exponential decay of the solution follows from the estimates proved
in lemma 2.1 provided that $b(t)=|f|$ decays exponentially fast. If
the weak solution exists globally and is unique, then a smooth
solution, if it exists globally, has to be equal to the weak
solution due to the uniqueness of the solution. Therefore, the weak
solution has to be smooth if a smooth solution exists.

The global existence of the weak solution was proved, for example,
in [5]-[7]. We give a slightly different proof. Let $D\subset
\mathbb{R}^3$ be a bounded domain. Denote by
$\{\phi_j\}_{j=1}^\infty$  the eigenvectors of the Stokes operator
$-P\Delta$ in $H=L^2(D)$, where $P$ is the Helmholtz-Leray projector
(see [1], [5], [6] or [7]). These eigenvectors are    
orthonormal in $H$,  and form a basis of $V$. They solve the problem:
 $$- P\Delta
\phi_j=\lambda_j\phi_j,\,\, \phi_j\in V;\quad 0<\lambda_1\le
\lambda_2 \dots, \lim_{j\to \infty}\lambda_j=\infty; \quad
((\phi_j,\phi_i))=\lambda_j\delta_{ij},$$ where $\delta_{ij}$ is the
Kronecker symbol. Let us look for a solution to (\ref{eq:3}) of the
form $v_m=\sum_{j=1}^m c_{jm}(t)\phi_j(x)$, where $ c_{jm}(t)$ are 
unknown functions. If $v_m$ is substituted in
equation (\ref{eq:3}) with $\eta=\phi_j$, then one gets:
\begin{equation}\label{eq:16}
c'_{jm}(t)+\nu\lambda_j c_{jm}(t)+ ((v_m\cdot \nabla)v_m,\phi_j)=
(f, \phi_j):=f_j(t), \,\, c_{jm}(0)=(v^0,\phi_j).
\end{equation}
Multiplying this equation by $c_{jm}$, summing up over $j$ from
$j=1$ to $j=m$,  taking into account that  
$$((v_m\cdot \nabla)v_m,v_m)=0, \quad \sum_{j=1}^m\lambda_j
c_{jm}^2=((v_m,v_m)):=G_m,$$ 
and denoting
$$g_m:=g_m(t):=\sum_{j=1}^mc_{jm}^2(t),$$ 
one gets 
\begin{equation}\label{eq:16a}
g'_m+2\nu \lambda_1 g_m\le g'_m+2\nu G_m \le
2|P_mf|g_m^{0.5},
\end{equation} 
where the inequality $\lambda_1 g_m\le G_m$ was used, $\lambda_1$ 
depends on $D$, 
and 
$$P_mf:=\sum_{j=1}^m f_j\phi_j, \quad \lim_{m\to
\infty}|P_mf-f|=0.$$ 
Inequality (\ref{eq:16a}) and Lemma 2.1 imply that 
\begin{equation}\label{eq:16c}
g_m(t)\le
ce^{-2\gamma t},
\end{equation}
 where $\gamma:=\nu \lambda_1$, the constant $c>0$ does 
not depend on $m$,
and it is assumed that $|f(t)|\le O(e^{-2\gamma t})$. 
The system (\ref{eq:16}) of ordinary
differential equations with the quadratic nonlinearity $$((v_m\cdot
\nabla)v_m,\phi_j)=\sum_{p,q=1}^m c_p(t)c_q(t)((\phi_p\cdot
\nabla)\phi_q,\phi_j)$$ has a local solution by the standard result.
Estimate (\ref{eq:16c}) shows that the local solution is bounded
uniformly with respect to $t$, and, consequently, the 
functions $c_{jm}(t)$, $1\le j \le m$,  exist
globally, that is, for all $t\ge 0$. Furthermore, there exists a 
subsequence, as $m\to \infty$, denoted $c_{jm}$
again, that converges weakly in $L^2([0,T))$ to a sequence
$\{c_j\}_{j=1}^\infty$, $c_j=c_j(t)$. From the estimate (\ref{eq:16c})
one concludes that
\begin{equation}\label{eq:16b}
g(t):=\sum_{j=1}^\infty c_{j}^2(t)\le ce^{-2\gamma t}.
\end{equation} 
Therefore, 
$c_j(t)=O(e^{-\gamma t})$ as $t\to \infty$. Moreover,
$\int_0^tG_m(s)ds$ is bounded uniformly with respect to $m$ and
$t\ge 0$. To prove this one uses an inequality similar to
(\ref{eq:4}): 
$$g_m+2\nu\int_0^tG_m(s)ds\le
g_m(0)+2\int_0^tb(s)g_m^{1/2}(s)ds,$$ 
and an estimate of $g_m^{1/2}$
similar to (\ref{eq:9}). 
 It follows from  (\ref{eq:16b}) that $\sum_{j=1}^\infty
c_j^2(t)\in L^\infty([0,T))$.
If the 
subsequence $c_{jm}$ converges weakly to $c_j$, then  
$v_m$ converges
weakly to a function $v$ in $W$. Let us check that the limiting function
$v=\sum_{j=1}^\infty c_{j}(t)\phi_j$ solves (\ref{eq:3}). Integrating
equation (\ref{eq:3}) with respect to $t$ one obtains
\begin{equation}\label{eq:17}
(v,\eta)+\nu \int_0^t((v,\eta))ds+
\int_0^t((v \cdot \nabla)v,\eta)ds=(v^0,\eta)+
\int_0^t(f,\eta)ds,\quad \forall \eta\in V.
\end{equation}
Let us compare (\ref{eq:17}) with the relation
\begin{equation}\label{eq:18}
(v_m,\eta)+\nu \int_0^t((v_m,\eta))ds+ \int_0^t((v_m \cdot \nabla)v_m,\eta)ds
=(v^0_m,\eta)+\int_0^t(P_mf,\eta)ds.
\end{equation}
Passing to the limit $m\to \infty$ in (\ref{eq:18}) yields
(\ref{eq:17}). The passage is straightforward in all the
terms, except for the term $\int_0^t((v_m \cdot \nabla)v_m,\eta)ds$.
This term can be rewritten as $-\int_0^t(v_m v_m, \nabla \eta)ds$.
The embedding operator from $V$ to $H$ is compact. Therefore
the weak convergence of $v_m$ in $L^2([0,T); V)$ implies the
convergence of the term $-\int_0^t(v_m v_m, \nabla \eta)ds$ to 
the integral $-\int_0^t(v v, \nabla \eta)ds=\int_0^t((v \cdot 
\nabla)v,\eta)ds$. Thus,  one
can pass to the limit in (\ref{eq:18}) and get (\ref{eq:17}). If
equation (\ref{eq:17}) holds, then one can differentiate  
(\ref{eq:17}) with respect to $t$ and obtain relation (\ref{eq:3}) 
for all $\eta\in V$. The set of the products  $\eta h_j(t)$, where 
$\eta\in V$ and the set $\{h_j(t)\}$
forms a basis of $L^2([0,T))$, 
is dense in the set $W$ in the norm of $L^2([0,T);V)$. Therefore,  
if relation  (\ref{eq:3})  holds for all $\eta\in V$ it holds
also for all $\eta\in W$. Consequently, the limiting function $v$
satisfies (\ref{eq:3}). 
The existence part of Theorem 1.2 is proved. \hfill $\Box$

\section{Unbounded domain}
Assume in this section that $D = \mathbb{R}^3$. Then inequality
(\ref{eq:7}) does not hold. We want to outline the proof of the
uniqueness result similar to Theorem 1.1 for unbounded domain
$\mathbb{R}^3$. Using  inequality (\ref{eq:Y}) one gets an analog of
inequality (\ref{eq:14a})
\begin{equation}\label{eq:20} ||u||_{L^4(\mathbb{R}^3)}^2 \leq
\nu H(t) + ch(t).
\end{equation}
This inequality and an inequality similar to (\ref{eq:14}) yield an
analog of inequality (\ref{eq:15}), and the uniqueness theorem
follows as in the case of a bounded domain $D$. This yields
\begin{theorem}\label{theorem3} If $D = \mathbb{R}^3$ then problem
(\ref{eq:3}) has at most one solution in $W_1$.
\end{theorem}

{\bf Acknowledgment.} The author thanks Dr. N. Pennington for a
discussion.


\end{document}